\documentclass[aps,pra,twocolumn,showpacs,superscriptaddress,floatfix]{revtex4-1}

\usepackage{amsmath}
\usepackage{amssymb}
\usepackage{bm}
\usepackage{graphicx}
\usepackage{float}
\usepackage{appendix}
\usepackage{longtable}
\usepackage{color}
\begin{document}
\title[]{Plasmon-assisted two-photon Rabi oscillations\\ in a semiconductor quantum dot -- metal nanoparticle heterodimer}
\author{Bintoro S. Nugroho}
\affiliation{Fakultas Matematika dan Ilmu Pengetahuan Alam, Universitas Tanjungpura, Jl. Jendral A. Yani, 78124 Pontianak, Indonesia}
\affiliation{Zernike Institute for Advanced Materials, University of Groningen, Nijenborgh 4, 9747 AG Groningen, The Netherlands}
\author{Alexander A. Iskandar}
\affiliation{Physics of Magnetism and Photonics Research Group, Institut Teknologi Bandung, Jl. Ganesa 10, 40132 Bandung, Indonesia}
\author{Victor A. Malyshev}
\affiliation{Zernike Institute for Advanced Materials, University of Groningen, Nijenborgh 4, 9747 AG Groningen, The Netherlands}
\author{Jasper Knoester}
\affiliation{Zernike Institute for Advanced Materials, University of Groningen, Nijenborgh 4, 9747 AG Groningen, The Netherlands}


\date{\today}

\begin{abstract}

Tho-photon Rabi oscillations hold potential for quantum computing and quantum information processing, because during a Rabi cycle a pair of entangled photons may be created. We theoretically investigate the onset of this phenomenon in a heterodimer comprising a semiconductor quantum dot strongly coupled to a metal nanoparticle. Two-photon Rabi oscillations in this system occur due to a coherent two-photon process involving the ground-to-biexciton transition in the quantum dot. The presence of a metal nanoparticle nearby the quantum dot results in a self-action of the quantum dot via the metal nanoparticle, because the polatization state of the latter depends on the quantum state of the former. The interparticle interaction gives rise to two principal effects:  (i) - enhancement of the external field amplitude and (ii) - renormalization of the quantum dot's resonance frequencies and relaxation rates of the off-diagonal density matrix elements, both depending on the populations of the quantum dot's levels. Here, we focus on the first effect, which results in interesting new features, in particular, in an increased number of Rabi cycles per  pulse as compared to an isolated quantum dot and subsequent growth of the number of entangled photon pairs per pulse.
We also discuss the destructive role of radiative decay of the excitonic states on two-photon Rabi oscillations for both an isolated quantum dot and a heterodimer.

\end{abstract}


\maketitle

\section{Introduction}
\label{Introduction}
%
%

Nanohybrids comprising a quantum emitter (QE) coupled to a metal nanoparticles (MNP) have been shown to exhibit a variety of novel optical properties which not only intriguing in their own right, but also offer great prospects for applications. In spite of considerable recent interest in hybrid QE-MNP systems, a number of questions regarding their optical properties have not been elucidated so far. One of those concerns the nonlinear optical interactions occurring in a QE-MNP nanohybrid, namely the onset of coherent two-photon processes, in particular, coherent two-photon absorption (TPA) and two-photon Rabi oscillations (TPRO). Although two-photon processes generally are weak, they may serve efficiently for various applications.
The characteristics of two-photon processes make them superior to other room-temperature schemes based on all-optical nonlinear processes~\cite{HayatSST2011}. The best known examples of practical applications of TPA are: microfabrication via 3D photopolymerization~\cite{MaruoOL1997,Baldacchini2016}, imaging~\cite{SvobodaNeuron2006}, and optical data storage~\cite{StricklerOptLett1991,CorredorAdvMat2006,MakarovJOSAB2007}. The principle of using TPA processes is based on the fact that many materials, while not being transparent for radiation in the visible, are transparent in the infrared. This allow one to reach the bulk materials with infrared light, where TPA processes may be used for optical applications such as the ones mentioned above.

The most interesting application of TPRO envisioned at the moment is the realization of a single-emitter source of pairs of polarization-entangled photons, which is of challenge for quantum computing and quantum information processing\cite{NeilsonBook2000,MacchiavelloBook2000}, as well as for quantum criptography~\cite{JenneweinPRL2000}.  A single semiconductor quantum dot (SQD) is considered an excellent candidate for efficient TPRO, as one may exploit the multiexciton states, in particular, biexciton states that naturally occur in these systems, to achieve this phenomenon.

During a cascade emission from the biexciton state of a SQD, a pair of polarization-entangled photons is created with its polarization determined by the spin of the intermediate exciton state~\cite{BensonPRL2000,StacePRB2003,FlissikowskiPRL2004,StevensonNature2006,AkimovPRL2006,AkopianPRL2006}. However, the entanglement is not complete because of the hyperfine splitting of one-exciton states forming the biexciton. This is the major obstacle for high quality polarization entanglement when using the cascade process. Additional manipulations are needed to reduce the splitting and improve entanglement~\cite{BensonPRL2000}. This problem is avoided by placing a SQD into a microcavity~\cite{delValleNJP2011,SchumacherOE2012} tuned exactly to the coherent two-photon resonance. Due to the biexciton shift, single-photon transitions are detuned and thus are effectively suppressed, while the coherent two-photon emission is Purcell-enhanced~\cite{delVallePRB2010}. An additional enhancement of the two-photon rate comes from the fact that the intermediate state denominator, determining the rate, is equal to the half of the biexciton binding energy which is usually on the order of several meV, i.e. relatively small for the second-order processes.

The TPRO represent a coherent process, involving the ground-to-biexciton transition, in which the intermediate one-exciton states play the role of
virtual states, not being populated, so that two photons created during one Rabi cycle are perfectly entangled by default. Importantly, this can be realized in a cavity-free configuration~\cite{BorriPRB2002,StuflerPRB2006}. Coherent control of the biexciton state of an isolated SQD has been achieved within a two-pulse~\cite{ChenPRL2002,FlissikowskiPRL2004} and a single-pulse~\cite{JayakumarPRL2013} scheme of excitation followed by the cascade emission of a  pair entangled photons. The first observation of the cavity-free TPRO of a single InGaAs SQD has been reported by Stufler {\it et al.}~\cite{StuflerPRB2006} together with a simplified theoretical approach explaining the peculiarities of the effect. A more comprehensive theory has been presented in Ref.~\onlinecite{MachnikowskiPRB2008}.

In this paper, we examine theoretically the TPRO in a SQD strongly coupled to a MNP, where the SQD, as in Ref.~\onlinecite{StuflerPRB2006}, is modelled as a ladder-like three-level system (ground, one-exciton, and bi-exciton states). It is well established that the presence of a MNP nearby a SQD has a vital influence on the optical response of the SQD as a consequence of the polarizability of the MNP. Notable phenomena that have been studied in detail are: bistable optical response~\cite{ArtusoNL2008,ArtusoPRB2010,MalyshevPRB2011,LiOE2012,NugrohoJCP2013}, linear and nonlinear Fano resonances~\cite{ZhangPRL2006,KosionisJPCC2012,NugrohoPRB2015}, gain without inversion~\cite{SadeghiNanotechnology2010}, and several other effects~\cite{SadeghiPRB2009,AntonPRB2012,NugrohoJOpt2017}. Our goal is to uncover the plasmonic effect on the TPRO of a single SQD.

This paper is organized as follows. In the next section, we present the model system and the mathematical formalism for its description. In Sec.~\ref{Results and discussion}, we report the results of numerical calculations of the TPRO for a set of parameters that is achievable in practice and discuss these.
In Sec.~\ref{Self-action effect}, we analyze under what conditions additional nonlinear effects, resulting from the self-action induced renormalization of the effective exciton energies and relaxation rates, may occur.
Section~\ref{Summary} summarizes the paper.

\section{Model and Formalism}
\label{Model and Formalism}
\begin{figure*}[ht]
\begin{center}
\includegraphics[width=0.7\linewidth]{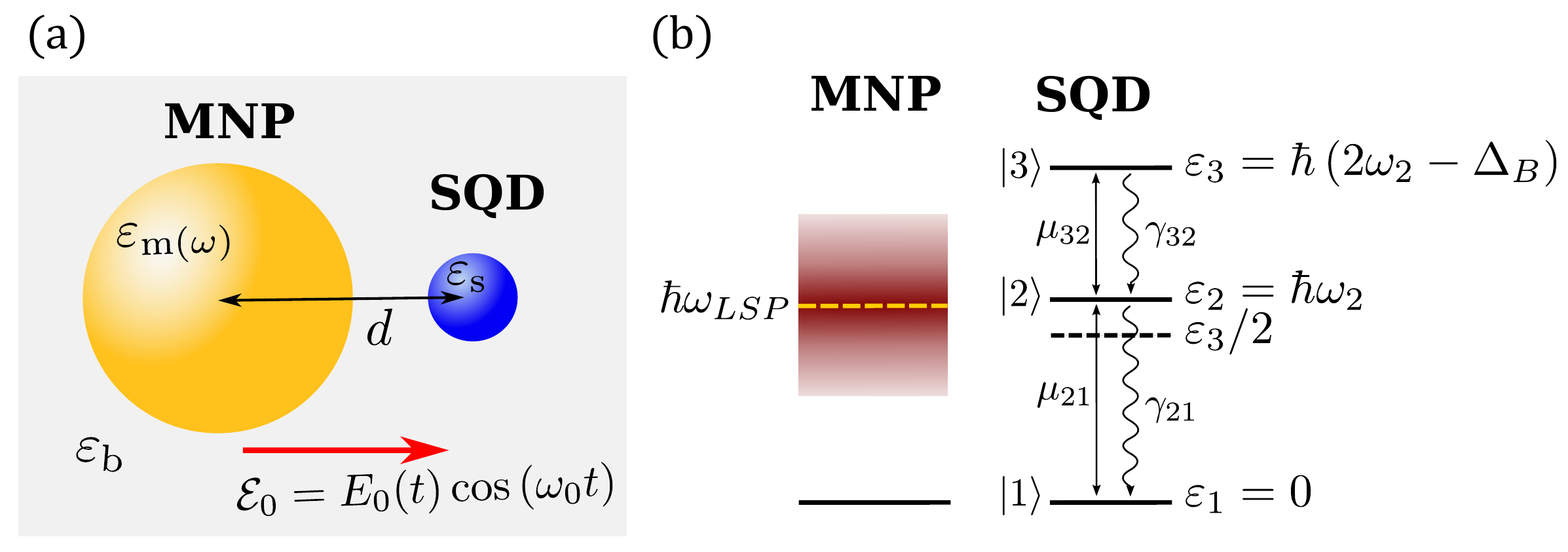}
\caption{\label{Schematics}(a)~Schematics of a SQD-MNP heterodimer subject to a pulsed applied field $\bm{\mathcal{E}}_0(t) = \boldsymbol{E}_0(t) \cos (\omega_0 t)$. The field is linearly polarized along the system axis (shown by the red arrow). $d$ is the SQD-MNP center-to-center distance, $r$ is the radius of the MNP, $\varepsilon_s$ and $\varepsilon_m(\omega)$ are the dielectric constant of the SQD and the MNP, respectively. The system is embedded in an isotropic and non-absorbing medium with permittivity $\varepsilon_b$. (b)~Energy diagrams of the MNP (left) and a ladder-type three-level SQD (right). The excited state of the MNP represents a broad line centered at the frequency of the LSP's resonance, $\omega_\mathrm{LSP}$, shown by the dashed yellow line. For the SQD, $|1 \rangle$, $|2 \rangle$, and $|3 \rangle$ are the ground, one-exciton, and bi-exciton  states, respectively. The energies of these states are $\varepsilon_1 =0$, $\varepsilon_2 =\hbar \omega_2$ and $\varepsilon_3 = 2\hbar(\omega_2 -\Delta_B/2)$, where $\hbar\Delta_B$ is the bi-exciton  binding energy. Allowed transitions with the corresponding transition dipole moments $\boldsymbol{\mu}_{21}$ and $\boldsymbol{\mu}_{32}$ are indicated by solid double-directed arrows. Wigging arrows denote the spontaneous decay with rates $\gamma_{32}$ and $\gamma_{21}$. The black dashed line shows the location of the coherent two-photon resonance $\omega_3/2 = \omega_2 - \Delta_B/2$ (with simultaneous absorption of two photons).
}
\end{center}
\end{figure*}

We theoretically investigate coherent light-matter interaction in a nanohybrid comprised of a SQD coupled to a closely spaced spherical MNP. The nanoparticles are embedded in an isotropic and lossless background with permittivity $\varepsilon_b$ and separated by a (center-to-center) distance $d$. The SQD is characterized by its (bulk) dispersionless dielectric constant $\varepsilon_s$, while the
MNP is described by the dielectric function $\varepsilon_m(\omega)$. The system is driven by a pulsed external field $\bm{\mathcal{E}}_0(t) = \boldsymbol{E}_0(t) \cos (\omega_0 t)$ with a carrier frequency $\omega_0$ and an amplitude $\boldsymbol{E}_0(t)$ which varies slowly on the scale of the optical period $2\pi/\omega_0$. The field is linearly polarized along the system's axis [see Fig.~\ref{Schematics}(a)]. The SQD and MNP radii as well as their center-to-center distance $d$ are assumed to be small compared to the optical wavelength, allowing us to apply the quasi-static approximation~\cite{Bohren1983,Maier2007} and to neglect the retardation in the SQD-MNP interaction.

Figure~\ref{Schematics}(b) shows the schematics of the electronic states and corresponding energy levels of the nanohybrid. The oscillating external field gives rise to oscillations of conducting electrons in the MNP, conventionally called localized surface plasmon (LSP). For the MNP's radii $r > 5$nm we consider, quantum-size effects are negligible~\cite{SchollNature2012}, so that for LSPs, the classical treatment can be safely applied~\citep{Maier2007}. Within this approach, the MNP's optical response can be described by its complex-valued frequency-dependent polarizability $\alpha \left(\omega\right)$, given by~\cite{Bohren1983,Maier2007}
\begin{equation}
  \alpha(\omega)= 4\pi r^3 \frac{\varepsilon_m(\omega)-\varepsilon_b}
                  {\varepsilon_m(\omega) + 2\varepsilon_b}~,
\label{alpha}
\end{equation}
The LSP's resonance frequency $\omega_\mathrm{LSP}$ is determined as the frequency at which the real part of $\alpha \left(\omega\right)$ is minimal (Fr\"{o}hlich condition).~\cite{Bohren1983} Thus, the electronic states of the LSP represent a ground state and a broad continuum of excited states as depicted in Fig.~1(b) (left).

The optical excitations in the SQD are excitons. In addition to the one-exciton states of the SQD, we incorporate also a bi-exciton state, corresponding to two excitations coupled by the Coulomb interaction.  In such a system, the degeneracy of the two one-exciton states is lifted due to the anisotropic electron-hole exchange, leading to two split linearly polarized one-exciton states
with a negligible energy splitting (on the order of a few tens of $\mathrm{\mu eV}$), one of which is dark, while the other is bright.
(see, e.g., Refs.~\onlinecite{StuflerPRB2006,JundtPRL2008,GerardotNJP2009}).
In this case, the ground state is coupled to the bi-exciton state via the linearly polarized one-exciton bright state. Thus, the system effectively acquires a three-level structure with a ground state $|1\rangle$, one- exciton state $|2\rangle$, and bi-exciton  state $|3\rangle$ with corresponding energies $0$, $\hbar\omega_{2}$, and $\hbar\omega_3 = 2\hbar(\omega_2 - \Delta_B/2)$, where $\hbar\Delta_B$ is the bi-exciton binding energy. Within this model, the allowed transitions, induced by the external field, are $|1\rangle \leftrightarrow |2\rangle$ and $|2\rangle \leftrightarrow |3\rangle$, which are characterized by the transition dipole moments $\boldsymbol{\mu}_{21} (= \boldsymbol{\mu}_{12})$ and $\boldsymbol{\mu}_{32} (= \boldsymbol{\mu}_{23})$, respectively (for the sake of simplicity, we assume that they are real). The states $|3\rangle$ and $|2\rangle$ spontaneously decay with rates $\gamma_{32}$ and $\gamma_{21}$[see Fig.~1(b)]. Note that the bi-exciton  state $|3\rangle$, having no allowed transition dipole moment from the ground state $|1\rangle$, can be reached either via consecutive $|1\rangle \rightarrow |2\rangle \rightarrow |3\rangle$ transitions or via the simultaneous absorption of two photons of frequency $\omega_3/2 = \omega_2 - \Delta_B/2$. At resonant excitation, each of these transitions can be addressed separately. This remains valid also for a pulsed excitation, if the pulses are spectrally narrower than the biexciton binding energy $\hbar\Delta_B$. We will predominantly focus our study on the resonant excitation of the coherent two-photon transition.

Excitons and plasmons, excited in the nanohybrid, interact with each other via the dipole-dipole interaction, which gives rise to a renormalization of the field experienced by both the SQD and MNP. The effects of the coupling can be inferred from the amplitude of the total field acting on the SQD, which equals the sum of the external field $\boldsymbol{E}_0(t)$ and the field produced by the MNP~\cite{ArtusoNL2008,ArtusoPRB2010,MalyshevPRB2011,LiOE2012,NugrohoJCP2013}.  With account for the contribution of higher multipoles, which is important if the MNP's radius $r$ is on the order of the SQD-MNP spacing $d$ (our case, see below), $\boldsymbol{E}_{SQD}(t)$ reads~\cite{YanPRB2008,NugrohoJOpt2017}
\begin{widetext}
\begin{equation}
\label{E_SQD}
  \boldsymbol{E}_{\mathrm{SQD}}(t) = \frac{1}{\varepsilon_s'}\left[ 1
                     + \frac{\alpha(\omega_0)}{2 \pi d^3} \right]\boldsymbol{E}_0(t)
                     + \frac{1}
                     {4 \pi^2 \varepsilon_0 \varepsilon_b \varepsilon_s'}
                     \sum_{n=1}^{\infty}\frac{(n+1)^2\alpha_n(\omega_0)}{d^{2n+4}}\boldsymbol{P}_{\mathrm{SQD}}(t)~,
\end{equation}
\end{widetext}
%
%
%
Here, $\varepsilon_s' = (\varepsilon_s + 2 \varepsilon_b)/(3\varepsilon_b)$ is the effective dielectric constant of the SQD, $\boldsymbol{P}_\mathrm{SQD}(t)$ is the amplitude of the dipole moment generated in the SQD [see below, Eq.~(\ref{P_SQD})]. Furthermore, the higher multipole polarizabilities $\alpha_n(\omega_0)$ are given by
\begin{equation}
\label{alpha_n}
  \alpha_n(\omega_0) = 4\pi r^{2n+1} \frac{\varepsilon_m(\omega_0)-\varepsilon_b}
                       {\varepsilon_m(\omega_0) + \frac{n+1}{n}\varepsilon_b} \ .
\end{equation}
The first term in Eq.~(\ref{E_SQD}) shows that the external field amplitude $\boldsymbol{E}_0(t)$ experiences renormalization due to the presence of the nearby MNP (second term in the square brackets). The last term in Eq.~(\ref{E_SQD}) reveals the electromagnetic self-action of the SQD via the MNP: the field acting upon the SQD depends on the amplitude of its own dipole moment $\boldsymbol{P}_\mathrm{SQD}(t)$. As will be shown below, this may considerably affects the hybrid's TPRO as compared to those of an isolated SQD.

It is worth noting that all polarizabilities in Eq.~(\ref{E_SQD}) are taken at the carrier frequency of the external field, $\omega_0$. Throughout this paper, we assume that the spectral width of the envelope $\boldsymbol{E}_0(t)$ is enough narrow to neglect variations of the polarizabilities over this interval. For the parameters chosen in our calculations, this condition is fulfilled easily.

We describe the optical process in the SQD by making use of the Lindblad quantum master equation for the density operator $\rho(t)$,
which in the rotating frame (with frequency $\omega_0$ of the external field) reads~\cite{Lindblad1976,Blum2012}
\begin{widetext}
\begin{subequations}
\label{MasterEqAndHamiltonian}
\begin{equation}
\label{DensityMasterEq}
\dot{\rho} = -\frac{i}{\hbar} \left[\mathcal{H}^{\mathrm{RWA}}(t),\rho\right ]
           + \mathcal{L}(\rho)
\end{equation}
\begin{equation}
\label{HamiltonianRWA}
H^\mathrm{RWA}(t) =  \hbar \left( \Delta_{21}|2 \rangle \langle 2| + \Delta_{31} |3 \rangle \langle 3| \right)
        -  \hbar \left[ {\Omega}_{21}(t) |2 \rangle \langle 1| + \Omega_{32}(t) |3 \rangle \langle 2| + H.c.\right ]~,
\end{equation}
\begin{equation}
\label{LindbladRWA}
\mathcal{L}(\rho) = \frac{\gamma_{21}}{2} \left( \left[ |1\rangle \langle 2| \rho,
           |2\rangle \langle 1| \right]  + \left[ |1\rangle \langle 2|,
           \rho\, |2\rangle \langle 1|\right]\right)
           + \frac{\gamma_{32}}{2} \left( \left[
           |2\rangle \langle 3|{\rho},|3\rangle \langle 2| \right]
           + \left[ |2\rangle \langle 3|,\rho\, |3\rangle \langle 2| \right]\right)~.
\end{equation}
\end{subequations}
\end{widetext}
Here, square brackets denote the commutator, $\mathcal{H}^{\mathrm{RWA}}(t)$ is the SQD Hamiltonian in the rotating wave approximation (RWA), and $\mathcal{L}(t)$ is the Lindblad operator describing the radiative relaxation in the system. In Eq.~(\ref{HamiltonianRWA}), $\hbar\Delta_{21} = \hbar(\omega_2 - \omega_0)$ and $\hbar(\Delta_{31} = \hbar(\omega_3 - 2\omega_0)$ are the energies of states $|2 \rangle$ and $|3 \rangle$ in the rotating frame, respectively. Alternatively, the former may be interpreted as the detuning away from the one-photon resonance and the latter as the detuning from the coherent two-photon resonance. $\Omega_{21}(t) = \boldsymbol{\mu}_{21}\cdot\boldsymbol{E}_{SQD}(t)/(2\hbar)$ and $\Omega_{32}(t) = \boldsymbol{\mu}_{32}\cdot \boldsymbol{E}_{SQD}(t)/(2\hbar)$ are the slowly varying Rabi amplitudes of $\boldsymbol{E}_{SQD}(t)$ for the corresponding transitions, where $\boldsymbol{E}_{SQD}(t)$ is the amplitude of the total field acting on the SQD. The latter is the sum of the applied field ${\boldsymbol{E}}_0(t)$ and the field produced by the plasmon oscillations in the MNP given by Eq.~(\ref{E_SQD})~\cite{ArtusoNL2008,ArtusoPRB2010,MalyshevPRB2011,LiOE2012,NugrohoJCP2013}. $\mathrm{H.c}$ stands for the Hermitian conjugate.

The SQD's dipole moment amplitude $\boldsymbol{P}_{\mathrm{SQD}}(t)$ in Eq.~(\ref{E_SQD}) is given by
\begin{equation}
\label{P_SQD}
  \boldsymbol{P}_{\mathrm{SQD}}(t) = \boldsymbol{\mu}_{21}\rho_{21}(t) + \boldsymbol{\mu}_{32}\rho_{32}(t)~.
\end{equation}
Thus, the Rabi amplitudes $\Omega_{21}(t)$ and $\Omega_{32}(t)$ may be expressed as follows:
\begin{subequations}
\label{Rabi frequencies}
\begin{equation}
        \Omega_{21}(t) = \widetilde{\Omega}^0_{21}(t) + G_1 \rho_{21}(t) + G_3 \rho_{32}(t) \ ,
\label{Omega21}
\end{equation}
\begin{equation}
        \Omega_{32}(t) = \widetilde{\Omega}^0_{32}(t) + G_3 \rho_{21}(t) + G_2 \rho_{32}(t) \ ,
\label{Omega32}
\end{equation}
\end{subequations}
where we introduced the quantities:
\begin{subequations}
\label{All_Omega}
\begin{equation}
        \widetilde{\Omega}^0_{21}(t) = \frac{1}{\varepsilon_s'}\left[ 1
                     +\frac{\alpha(\omega_0)}{2 \pi d^3} \right]\Omega^0_{21}(t) \ ,
\label{Omega021}
\end{equation}
\begin{equation}
        \widetilde{\Omega}^0_{32}(t) = \frac{1}{\varepsilon_s'}\left[ 1
                     +\frac{\alpha(\omega_0)}{2 \pi d^3} \right]\Omega^0_{32}(t) \ ,
\label{Omega032}
\end{equation}
\end{subequations}
and
\begin{subequations}
\label{All_G}
\begin{equation}
        G_1 = \frac{\boldsymbol{\mu}_{21}\cdot\boldsymbol{\mu}_{21}}{16 \pi^2 \hbar \varepsilon_0 \varepsilon_b \varepsilon_s'}\sum_n \frac{(n+1)^2\alpha_n(\omega_0)}
                     {d^{2n+4}} \ ,
\label{G1}
\end{equation}
\begin{equation}
        G_2 = \frac{\boldsymbol{\mu}_{32}\cdot\boldsymbol{\mu}_{32}}
                     {16 \pi^2 \hbar \varepsilon_0 \varepsilon_b \varepsilon_s'}
                     \sum_n \frac{(n+1)^2\alpha_n(\omega_0)}
                     {d^{2n+4}}\ ,
\label{G2}
\end{equation}
\begin{equation}
        G_3 = \frac{\boldsymbol{\mu}_{21}\cdot\boldsymbol{\mu}_{32}}
                     {16 \pi^2 \hbar \varepsilon_0 \varepsilon_b \varepsilon_s'}
                     \sum_n \frac{(n+1)^2\alpha_n(\omega_0)}
                     {d^{2n+4}}\ .
\label{G3}
\end{equation}
\end{subequations}
Here, $\Omega^0_{21}(t) = \boldsymbol{\mu}_{21} \cdot \boldsymbol{E}_0(t)/(2\hbar)$ and $\Omega^0_{32}(t) = \boldsymbol{\mu}_{32} \cdot \boldsymbol{E}_0(t)/(2\hbar)$ are the Rabi amplitudes of the external field for the corresponding transitions. As is seen, the external field experiences a renormalization due to the presence of a nearby MNP, which is reflected in the factor $1 + \alpha(\omega_0)/(2\pi d^3)$.
The complex-valued quantities $G_1 = G_1^R + iG_1^I$, $G_2 = G_2^R + iG_2^I$, and $G_3 = G_3^R + iG_3^I$ represent the so-called feedback parameters, describing the self-interaction of the SQD via the MNP~\cite{ArtusoNL2008,ArtusoPRB2010,MalyshevPRB2011,LiOE2012,NugrohoJCP2013}. They contain all details of the SQD-MNP coupling, such as the contribution of higher multipoles, material properties, and geometry of the system.

Using the above, the set of equations for the density matrix elements $\rho_{ij}$ \, ($i,j = 1,2,3$), governing the optical dynamics of the SQD in the presence of the nearby MNP, reads~\cite{NugrohoJOpt2017}
\begin{widetext}
\begin{subequations}
\label{all_dR1}
\begin{equation}
\dot{\rho}_{11} = \gamma_{21} \rho_{22} + i(\Omega_{21}^* \rho_{21} - \Omega_{21} \rho^*_{21})~,
\label{rho11}
\end{equation}
%
\begin{equation}
\dot{\rho}_{22} = - \gamma_{21} \rho_{22} + \gamma_{32} \rho_{33}
                + i (\Omega_{21} \rho^*_{21} - \Omega^*_{21} \rho_{21}
                + \Omega^*_{32} \rho_{32} - \Omega_{32} \rho^*_{32})~,
\label{rho22}
\end{equation}
\begin{equation}
\dot{\rho}_{33} = -\gamma_{32} \rho_{33} + i (\Omega_{32} \rho^*_{32} - \Omega^*_{32} \rho_{32})~,
\label{rho33}
\end{equation}
\begin{equation}
\dot{\rho}_{21} = -\left[ i \Delta_{21} + \frac{1}{2}\gamma_{21} \right] \rho_{21}
                + i ( \Omega^*_{32} \rho_{31} - \Omega_{21} Z_{21} )~,
\label{rho21}
\end{equation}
\begin{equation}
\dot{\rho}_{32} = -\left[ i\Delta_{32} + \frac{1}{2} (\gamma_{32} + \gamma_{21}) \right] \rho_{32}
                - i ( \Omega^*_{21} \rho_{31} + \Omega_{32} Z_{32} )~,
\label{rho32}
\end{equation}
\begin{equation}
\dot{\rho}_{31} = -\left[ i\Delta_{31} + \frac{1}{2}\gamma_{32} \right ] \rho_{31}
                + i (\Omega_{32} \rho_{21} - \Omega_{21} \rho_{32})~,
\label{rho31}
\end{equation}
\end{subequations}
\end{widetext}
where $\Delta_{32} = \omega_3 - \omega_2 - \omega_0$ is the detuning away from the $|3 \rangle \leftrightarrow |2 \rangle$ transition and $Z_{ji} = \rho_{jj}-\rho_{ii}$ denotes the population difference between the states $|j\rangle$ and  $|i\rangle$. Here, we suppressed the time dependence of all dynamic variables.

\section{Results and discussion}
\label{Results and discussion}
\subsection{Isolated SQD}
\label{Isolated SQD}
First, we consider the TPRO of an isolated SQD. This will serve as reference case for the hybrid analyzed in Sec.~\ref{Hybrid SQD}. The new aspect, added in this section as compared to previous studies of the isolated SQD~\cite{StuflerPRB2006,MachnikowskiPRB2008}, is an analysis of the destructive effect of the exciton's spontaneous decay on the TPRO.
For the SQD, we use the set of parameters of a CdSe/ZnSe quantum dot with radius 3 nm~\cite{FlissikowskiPRL2004}:
the energies of the bare one-exciton and bi-exciton transitions are $\hbar \omega_2 = 2.36~\mathrm{eV}$ and $\hbar \omega_3 = \hbar (2\omega_2 - \Delta_B)$ with $\hbar\Delta_B = 20~\mathrm{meV}$ ($\Delta_B \approx 30\,\mathrm{ps}^{-1}$), the population radiative decay rates $\gamma_{21} = 1/220~\mathrm{ps^{-1}}$ and $\gamma_{32} = 1/120~\mathrm{ps^{-1}}$. From these data, the transition dipole moments are evaluated as $\mu_{21} = 0.6~e\cdot\mathrm{nm}$ and $\mu_{32} = 0.8~e\cdot\mathrm{nm}$. For the SQD's dielectric constant $\varepsilon_s$, the typical value $\varepsilon_s = 6$ is taken.
In the following, we denote $\Omega_0(t) \equiv \Omega^0_{21}(t)$ and $\Omega(t) \equiv \Omega_{21}(t)$. Accordingly, $\Omega^0_{32}(t) = (\mu_{32}/\mu_{21}) \Omega_0(t)$ and $\Omega_{32}(t) = (\mu_{32}/\mu_{21}) \Omega(t)$.

The system is subjected to a resonant pulsed external field with Gaussian Rabi amplitude
\begin{equation}
\label{Pulse}
    \Omega_0(t) = \frac{A}{\sqrt{\pi}t_0} \exp\left[-\left(\frac{t - t_d}{t_0}\right)^2\right]~,
\end{equation}
tuned into the two-photon transition ($\omega_0 = \omega_3/2, \Delta_{21} = \Delta_B/2$). In Eq.~(\ref{Pulse}), $A = \int_{-\infty}^{\infty} \Omega_0(t)dt$ is the pulse area, $t_d$ is the delay time until the pulse maximum, and $t_0$ is the parameter determining the pulse duration $t_p$ which we define as the pulse full width at half maximum, $t_p = 2\sqrt{\ln 2}\,t_0$. We are primarily interested in the behavior of the biexciton state population $\rho_{33}$.

\subsubsection{Adiabatic limit}
\label{AdiabaticLimit}
Here, we recall the principal results of the adiabatic theory of the TPRO of an isolated SQD developed in Refs.~\onlinecite{StuflerPRB2006,MachnikowskiPRB2008}. Adiabaticity means that the coherently driven evolution of the system is slow enough, such that at each instant of time, the system follows the adiabatic instantaneous eigenstate of the Hamiltonian~(\ref{HamiltonianRWA})~(see, e.g. the textbook~\onlinecite{MessiahBook1966}).
This limit involves certain relationships between the system's parameters $\Omega_0(t)$, $t_0$, $\Delta_B$, and $\gamma_{32}^{-1}$.
First, adiabaticity requires the inequality $\Delta_Bt_0 \gg 1$ which means that the spectral width of the incident pulse, $\approx~1/(2t_0)$, is much smaller than the detuning away from the one-exciton resonance, $\Delta_B/2$, so that the transition to the latter is almost forbidden. This criterion is obeyed better the longer the pulse is. On the other hand, the pulse duration must be much shorter than the biexciton decay time, $t_0 \ll \gamma_{32}^{-1}$. Otherwise, the one-exciton state will be incoherently populated via the biexciton state during the pulse action, thus destroying the coherent TPRO.

We restrict ourselves to a simplified version of the adiabatic theory of the TPRO, assuming that the Rabi amplitude $A/(\sqrt{\pi} t_0)$ is much smaller than half the biexciton binding frequency $\Delta_B/2$. In this limit, the perturbation theory can be applied. A more comprehensive treatment (free of this assumption) can be found in Refs.~\onlinecite{StuflerPRB2006,MachnikowskiPRB2008}.

The two-photon Rabi amplitude $\Omega_2(t)$, calculated using second-order perturbation theory, reads:
\begin{equation}
\label{Omega2}
    \Omega_2(t) = \frac{2\Omega_{21}(t)\Omega_{32}(t)}{\Delta_B\varepsilon_s'^2} = \frac{2}{\Delta_B}\frac{\mu_{32}}{\mu_{21}}\left[\frac{\Omega_0(t)}{\varepsilon_s'}\right]^2~.
\end{equation}
Accordingly, the area of the "two-photon" pulse can be written as
\begin{equation}
\label{A2}
    A_2 = \int_{-\infty}^{\infty} \Omega_2(t) dt = \sqrt{\frac{2}{\pi}}\frac{\mu_{32}}{\mu_{21}}\left(\frac{A}{\varepsilon_s'}\right)^2 \frac{1}{\Delta_B t_0}~.
\end{equation}
The $A_2$-dependence of the biexciton population is given by~\cite{StuflerPRB2006,MachnikowskiPRB2008}
\begin{equation}
\label{rho33-vs-A2}
    \rho_{33} = \sin^2 \frac{A_2}{2}~.
\end{equation}
Thus, the biexciton population acquires its first maximum, $\rho_{33} = 1$ (full population), at $A_2 = \pi$, similar to that of the one-photon counterpart, except that the area of the "two-photon" pulse, $A_2$, depends quadratically on the area of the incident pulse, $A$ [see Eq.~(\ref{A2}].

Equating $A_2 = \pi$, one can evaluate the incident pulse area $A$ at which the biexciton state is fully populated:
\begin{equation}
\label{A(A2 = pi)}
    A_{A_2 = \pi}  = \varepsilon_s'\left(\pi\sqrt\frac{\pi}{2}\frac{\mu_{21}}{\mu_{32}}\Delta_B t_0 \right)^{1/2}~.
\end{equation}
From this estimate it follows that on increasing the incident pulse duration $t_0$, a larger area of the incident pulse is needed to invert the system. Moreover, it increases proportionally to $t_0^{1/2}$, which is consistent with our numerical calculations (see Sec.~\ref{Results}). Note that corresponding Rabi amplitude of the incident pulse $\Omega_0(t)$ decreases inversely proportional to $t_0^{1/2}$. The latter, together with narrowing of the pulse spectrum for increasing $t_0$, points on the fact that the contribution of the intermediate one-exciton state to the TPRO will be less important for pulses of a larger duration, which is also in agreement with numerical data (see Sec.~\ref{Results}).

\subsubsection{Numerical results}
\label{Results}
\begin{figure}
\begin{center}
\includegraphics[width=\columnwidth]{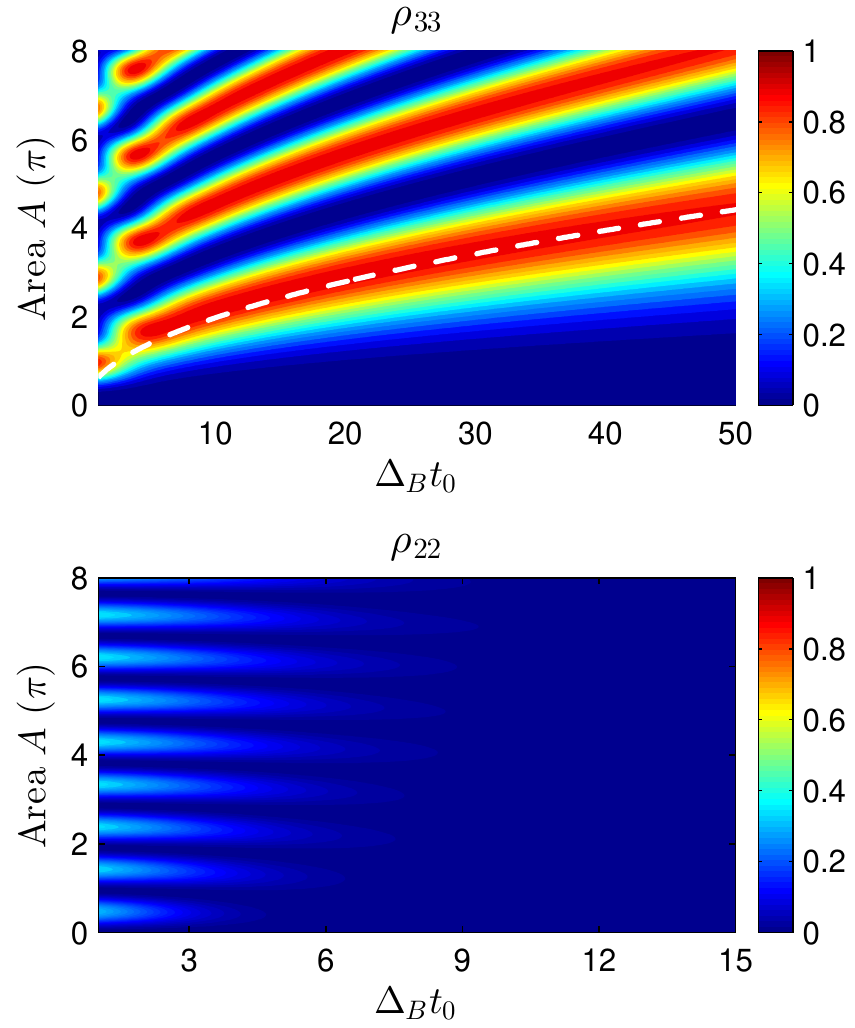}
\end{center}
\caption{\label{ContourPlotsRho33andRho22Single} Contour plots of the TPRO of an isolated SQD. Upper panel - population of the biexciton state, $\rho_{33}$. The white-dashed curve shows the $t_0$-dependence of the incident pulse area A, at which $\rho_{33}$ acquires its first maximum, plotted according to Eq.~(\ref{A(A2 = pi)}) with a correction coefficient 0.62. Lower panel - population of the one-exciton state, $\rho_{22}$.}
\end{figure}
\begin{figure}
\begin{center}
\includegraphics[width=\columnwidth]{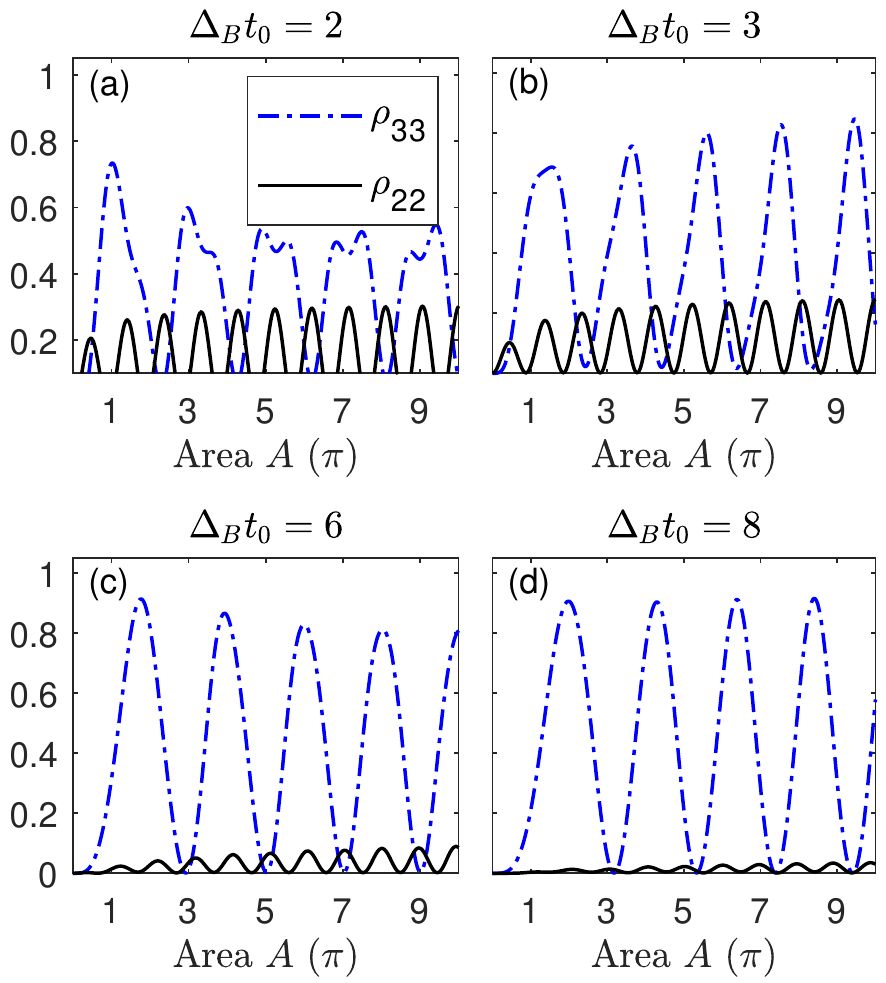}
\end{center}
\caption{\label{AreaDependenceRho33andRho22Single} Area dependence of the populations of the bi- and one-exciton states for different incident pulse durations indicated above each panel. $\Delta_B = 30\,\mathrm{ps}^{-1}$.}
\end{figure}
\begin{figure}
\begin{center}
\includegraphics[width=\columnwidth]{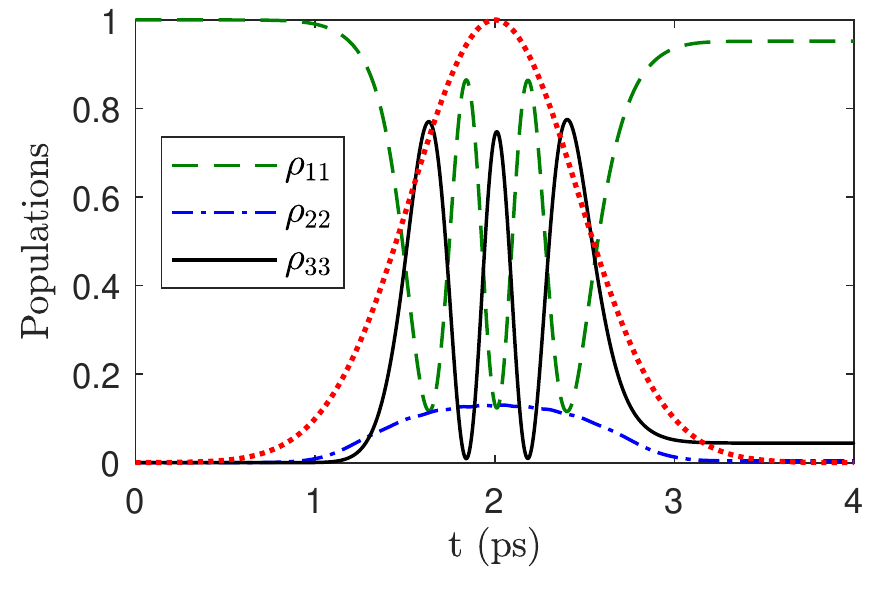}
\end{center}
\caption{\label{PopulationsDynamicsShortPulseSingle}Population dynamics calculated for the incident pulse with area $A = 9\pi$, duration $t_0 = 20/\Delta_B = 2/3$~ps, and delay time $t_d = 2$ ps. The red-dotted curve shows the profile of the incident Gaussian pulse, $\exp\{-[(t - t_0)/t_d]^2\}$.}
\end{figure}
In Fig.~\ref{ContourPlotsRho33andRho22Single}, we present the contour plots of the populations of the bi- and one-exciton states obtained by numerically integrating Eqs.~(\ref{rho11})~-~(\ref{rho31}). The figure clearly exhibits TPRO when following the behavior of $\rho_{33}$ as a function of pulse area at constant pulse duration. In the $\rho_{33}$ plot, the white-dashed curve displays Eq.~(\ref{A(A2 = pi)}) with a correction prefactor 0.62 in its right-hand side. As is seen, the theoretical prediction, based on the adiabatic theory, gives an excellent description for large values of $\Delta_Bt_0$ and breaks down for $\Delta_Bt_0$ on the order of unity, as expected.

Figure~\ref{AreaDependenceRho33andRho22Single} shows the area dependence of $\rho_{33}$ and $\rho_{22}$ for the region of failure of the adiabatic theory ($\Delta_B t_0 \sim 1$). As follows from the plots, this limit describes the actual behavior better for increasing $t_0$ and already starting at  $t_0 = 8/\Delta_B$, the adiabatic approcimation can be safely applied. The slight increase of the one-exciton population $\rho_{22}$  we relate to the direct excitation of this state due to its finite width $\gamma_{21}^{-1}$, as well as its population via the biexciton state, both occurring during each Rabi cycle. The lower plot in Fig.~2 presents a more detailed picture of the $\rho_{22}$ contribution to the TPRO.
\begin{figure}
\begin{center}
\includegraphics[width=\columnwidth]{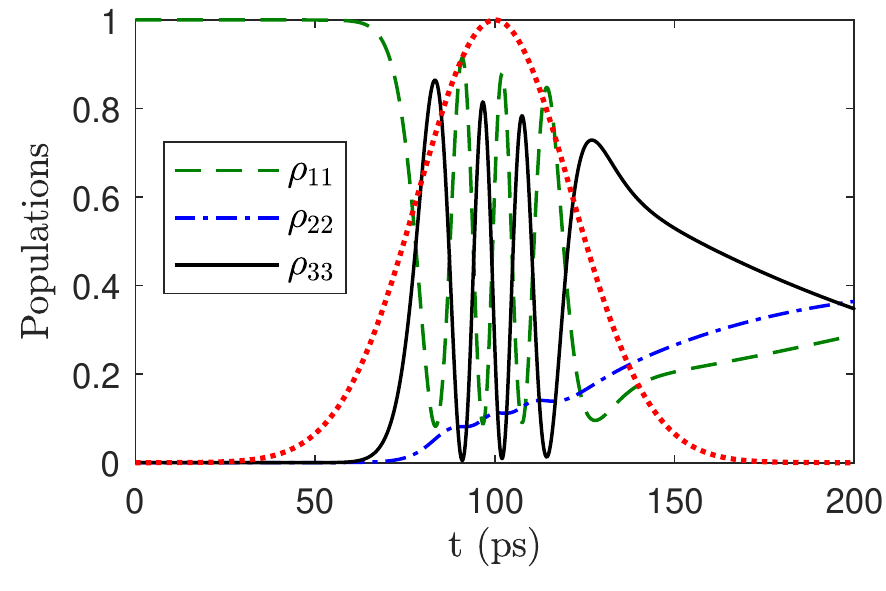}
\end{center}
\caption{\label{PopulationsDynamicsLongPulseSingle} Same as in Fig.~\ref{PopulationsDynamicsShortPulseSingle}, but now for $A = 50\pi$, $t_0 = 900/\Delta_B = 30$ ps, and $t_d = 100$ ps.}
\end{figure}

An example of the time evolution of populations for a short pulse ($t_0 = 20/\Delta_B = 2/3~\mathrm{ps} \ll \gamma_{21}, \gamma_{32})$ is presented in Fig.~\ref{PopulationsDynamicsShortPulseSingle}. Here, we are almost in the purely adiabatic regime. Indeed, the pulse spectrum, having a width that may be estimated as $1/(2t_0)$, is 20 times narrower than half the biexciton binding frequency $\Delta_B/2$ (detuning away from the one-exciton resonance), the condition for adiabaticity~\cite{StuflerPRB2006,MachnikowskiPRB2008}. On the other hand, the maximal value of the Rabi amplitude of the incident pulse, $A/(\sqrt{\pi}t_0)$, for the value $A = 9\pi$ used to calculate this figure is equal to $0.8 \Delta_B$, i.e. even larger than $\Delta_B/2$. This explains the relatively elevated population of the one-exciton state during the pulse action. Effects of radiative decay of the bi- and one-exciton states, giving rise to the incoherent population of the one-exciton state, are negligible in the considered case, because the duration of the incident pulse is much shorter than the spontaneous decay time of the biexciton state, $t_0 = 2/3\, \mathrm{ps} \ll \gamma_{32}^{-1} = 120\, \mathrm{ps}$.

The results presented in Figs.~\ref{ContourPlotsRho33andRho22Single}~-~\ref{PopulationsDynamicsShortPulseSingle}  were obtained for relatively short incident pulses, $t_0 \leq 50/\Delta_B = 5/3$ ps. At such pulse durations, the effects of the spontaneous decay are marginal, as was already noticed. It is of interest to look at longer pulses, when this process comes into play. Figure~\ref{PopulationsDynamicsLongPulseSingle} shows the population dynamics calculated for a significantly longer incident pulse, $t_0 = 900/\Delta_B = 30$ ps. Here, the pulse spectrum width $\approx 1/(2t_0)$ is  900 times narrower than half the biexciton binding frequency, $\Delta_B/2$. From this point of view, we are deep in the adiabatic limit. The maximal value of the Rabi amplitude of the incident pulse, $A/(\sqrt{\pi}t_0)$, for $A = 50\pi$ used in this case is equal to $0.1 \Delta_B$, i.e. significantly smaller than $\Delta_B/2$. Thus, contrary to the previous case of a short pulse, the one-exciton state would not be expected to be noticeably excited. Nevertheless, as is seen in Fig.~\ref{PopulationsDynamicsLongPulseSingle}, its population progressively grows during the pulse action and even onward, not disappearing after the pulse has passed, as it happens in the adiabatic case. We explain this behavior from the fact that the time scale of the incident pulse, $t_0 = 30$~ps (duration $\approx 60$ ps), is already comparable with the time scale of spontaneous decay of the biexciton state, $\gamma_{32}^{-1} = 120$ ps, Thus, the latter process leads to incoherent population of the one-exciton state during the pulse action and further and thereby destroys the TPRO. Thus, a tradeoff exists between the adiabaticity, supporting the pure TPRO, and the spontaneous decay which brakes down the coherence of the TPRO.

\subsection{SQD-MNP hybrid}
\label{Hybrid SQD}
\begin{figure}[ht!]
\begin{center}
\includegraphics[width=\columnwidth]{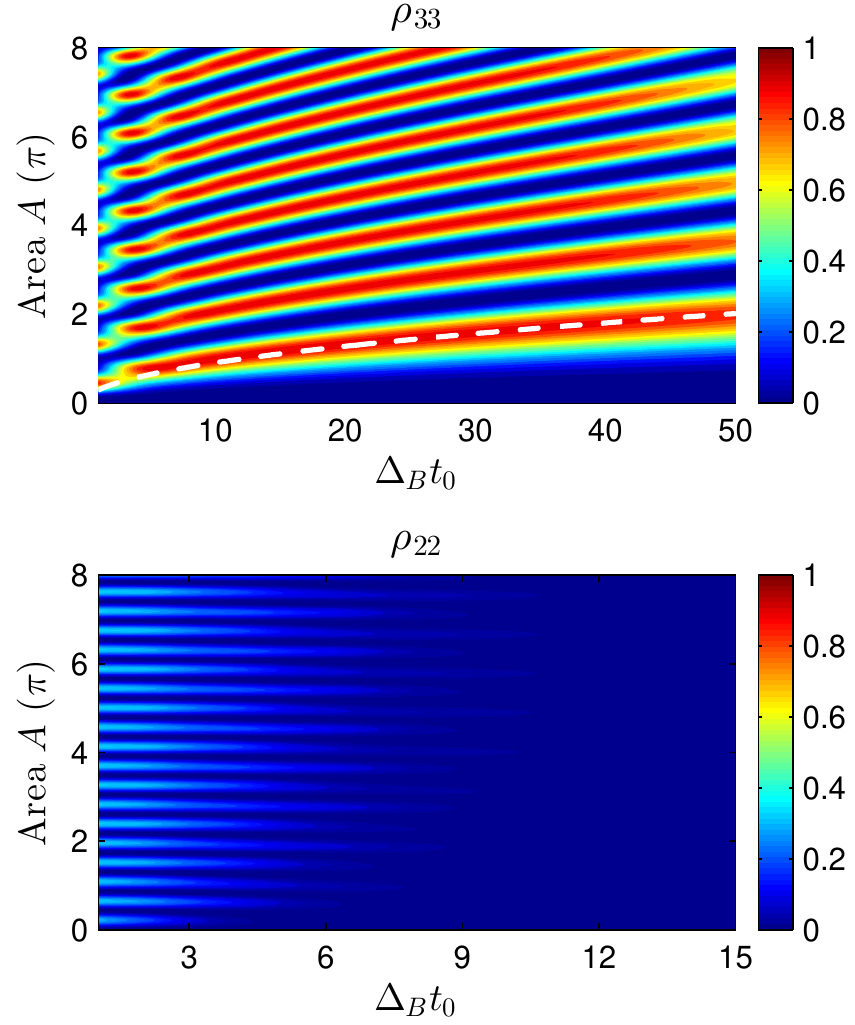}
\end{center}
\caption{\label{ContourPlotsRho33andRho22Hybrid} Contour plots of the TPRO of a SQD-MNP hybrid calculated for a center-to-center distance $d = 18$ nm. Upper panel - population of the biexciton state, $\rho_{33}$. The dashed-white curve shows the $t_0$-dependence of the incident pulse area $\widetilde{A}$, at which $\rho_{33}$ acquires its first maximum, plotted according to Eq.~(\ref{tildeA(A2 = pi)}) with a correction prefactor 0.62. Lower panel - population of the one-exciton state, $\rho_{22}$.}
\end{figure}

In this section, we analyse the TPRO of a hybrid comprizing a CdSe/ZnSe quantum dot and a nearby spherical gold MNP both embedded in a host with  permittivity $\varepsilon_b = 2.16$ (silica). For the SQD, we keep the same parameters as in Sec.~\ref{Isolated SQD}. The gold nanosphere is chosen to have radius $a = 12~\mathrm{nm}$. Its dielectric function $\varepsilon_m$ is calculated by making use of an improved Drude-like model~\cite{DerkachovaPlasmonics2016}. The corresponding surface plasmon resonance is found to be $\hbar\omega_{sp} = 2.34$ eV. The SQD-MNP separation $d$ is chosen small enough to get a strong coupling between the nanoparticles. We perform calculations similar to those presented in Sec.~\ref{Isolated SQD} to explicitly see the effect of the presence of a nearby MNP.

Figure~\ref{ContourPlotsRho33andRho22Hybrid} shows the contour plots for the populations of the bi- and one-exciton states, calculated for the SQD-MNP center-to-center distance $d = 18$~nm. In the $\rho_{33}$ plot, the white-dashed curve displays the dependence of the incident pulse area $\widetilde{A}$ at which $\rho_{33}$ attains its first maximum, calculated within the adiabatic limit adjusted to a hybrid. Here, Eq.~(\ref{A(A2 = pi)}) for an isolated SQD has to be modified by replacing the Rabi amplitude $\Omega_0(t)$ of the external field by the renormalized quantity $\widetilde{\Omega}_0(t) = [1 + \alpha(\omega_0)/(2\pi d^3)] \Omega_0(t)$ [see Eqs.~(\ref{Omega21}) and~(\ref{Omega32})]. Then for $\widetilde{A}_(A_2 = \pi)$ one gets
\begin{equation}
\label{tildeA(A2 = pi)}
    \widetilde{A}_{A_2 = \pi}  = \varepsilon_s' \left| 1 + \frac{\alpha(\omega_0)}{2 \pi d^3} \right|^{-1}
    \left(\pi\sqrt\frac{\pi}{2}\frac{\mu_{21}}{\mu_{32}}\Delta_B t_0 \right)^{1/2}~.
\end{equation}
In deriving Eq.~(\ref{tildeA(A2 = pi)}), we only took into account the enhancement effect and neglected the self-action of the SQD, which is described by the feedback constants $G_i$, Eqs.~(\ref{G1})~-~(\ref{G3}). For the system parameters used here, the self-action effect is negligible (see Sec.~\ref{Self-action effect}).

In order to fit the numerical data, Eq.~(\ref{tildeA(A2 = pi)}) should be corrected by a prefactor 0.62 which does not depend on the SQD-MNP spacing $d$. As is seen, modified in this way Eq.~(\ref{tildeA(A2 = pi)}) works excellently for large values of $\Delta_Bt_0$, while breaking down for values of $\Delta_Bt_0$ on the order of unity, exactly as in the case of the isolated SQD.

We note that in the case of the hybrid, more Rabi cycles occur at the same area $A$ of the incident pulse than for the isolated SQD (compare Figs.~\ref{ContourPlotsRho33andRho22Hybrid} and~\ref{ContourPlotsRho33andRho22Single}). The origin of this difference lies in the enhancement of the external field magnitude in the presence of a nearby MNP by the factor $|1 + \alpha(\omega_0)/(2\pi d^3)|$. This feature will be present in all figures following below.
For the hybrid parameters used here, this factor is $\approx 2.2$, which explains the doubling of the number of Rabi cycles for the hybrid in comparison with the isolated SQD, as observed from Figs.~\ref{ContourPlotsRho33andRho22Hybrid} and~\ref{ContourPlotsRho33andRho22Single}.

\begin{figure}
\begin{center}
\includegraphics[width=\columnwidth]{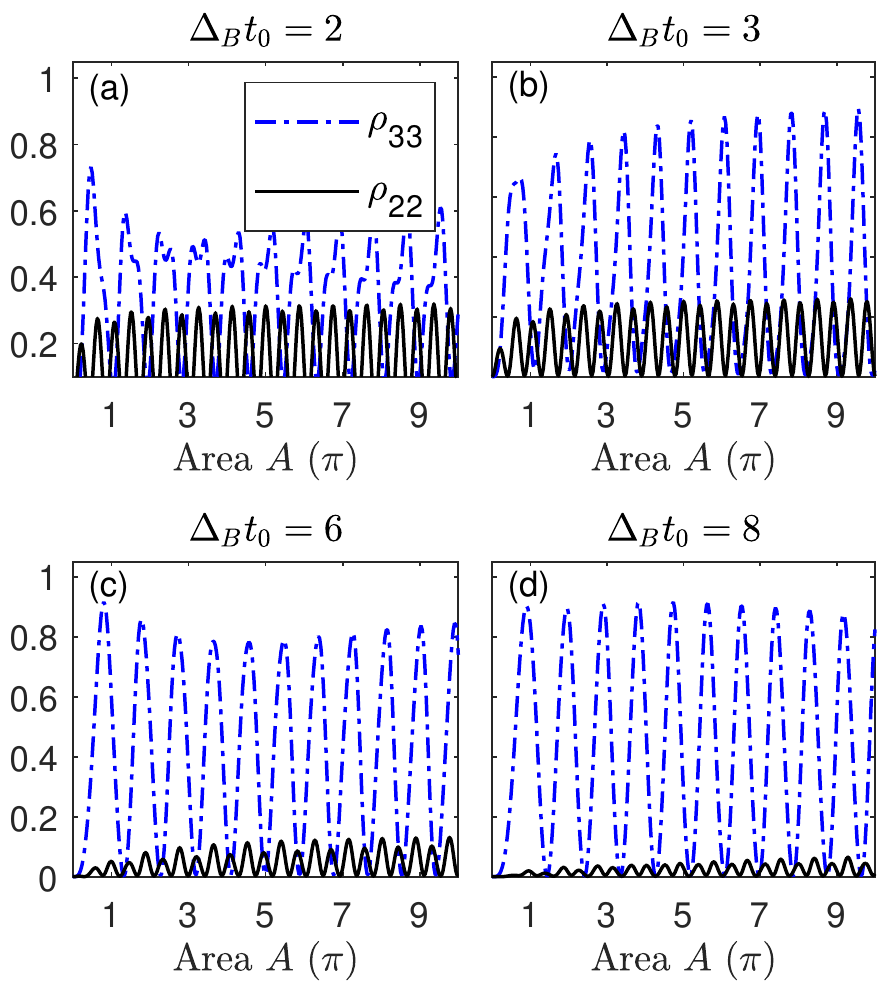}
\end{center}
\caption{\label{AreaDependenceRho33andRho22Hybrid} Same as in Fig.~\ref{AreaDependenceRho33andRho22Single}, but now for a hybrid with center-to-center distance $d =18$ nm.}
\end{figure}

In Fig.~\ref{AreaDependenceRho33andRho22Hybrid}, we plotted the area dependence of $\rho_{33}$ and $\rho_{22}$ for the region of failure of the adiabatic theory. As is seen, the adiabatic limit approximates the actual behavior better for growing $t_0$ and already starting from $t_0 = 8/\Delta_B$ can be safely applied. The slight buildup of the one-exciton population $\rho_{22}$ on increasing the pulse area has the same origin as in the case of an isolated SQD (see Sec.~\ref{Results}).

\begin{figure}
\begin{center}
\includegraphics[width=\columnwidth]{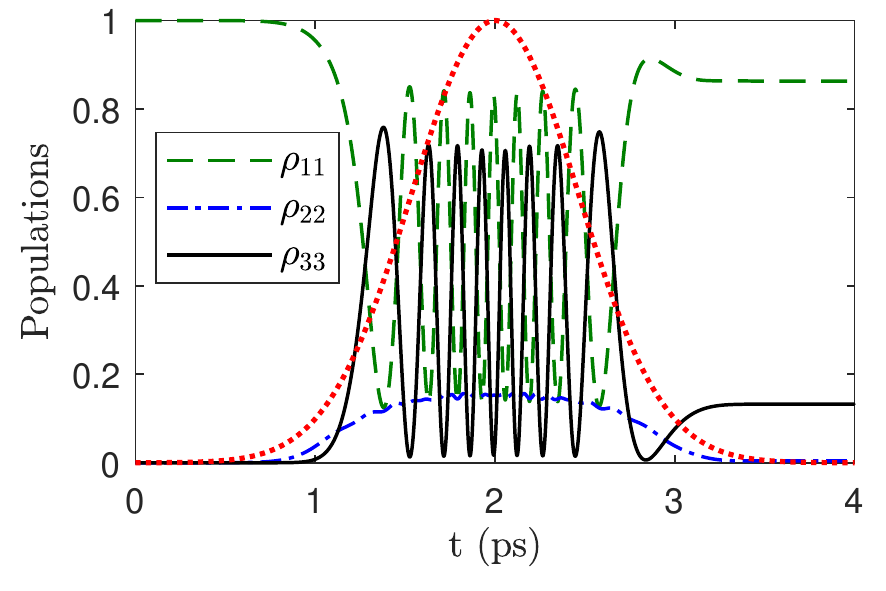}
\end{center}
\caption{\label{PopulationsDynamicsShortPulseHybrid} Same as in Fig.~\ref{PopulationsDynamicsShortPulseSingle}, but now for a hybrid with center-to-center distance $d =18$ nm.}
\end{figure}

\begin{figure}
\begin{center}
\includegraphics[width=\columnwidth]{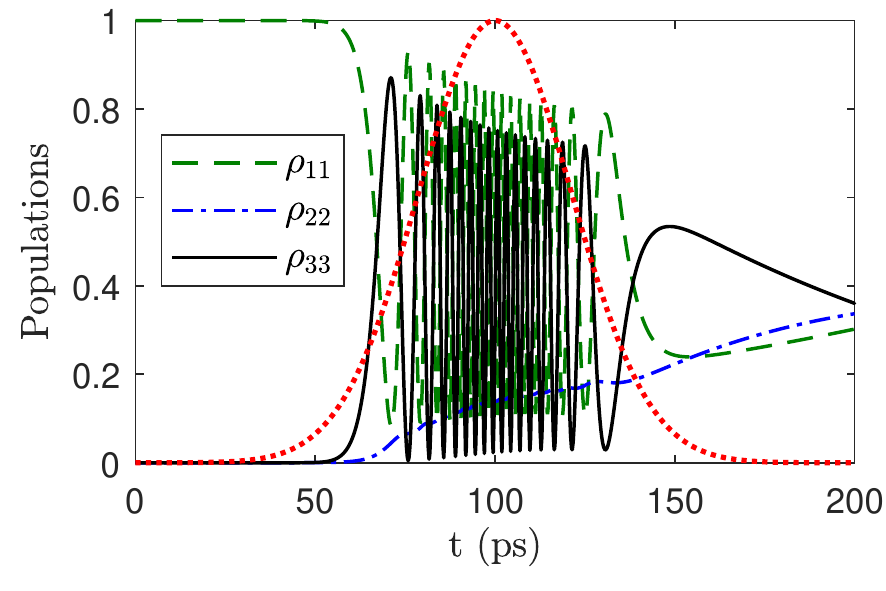}
\end{center}
\caption{\label{PopulationsDynamicsLongPulseHybrid} Same as in Fig.~\ref{PopulationsDynamicsLongPulseSingle}, but now for a hybrid with center-to-center distance $d =18$ nm.}
\end{figure}
\begin{figure}
\begin{center}
\includegraphics[width=\columnwidth]{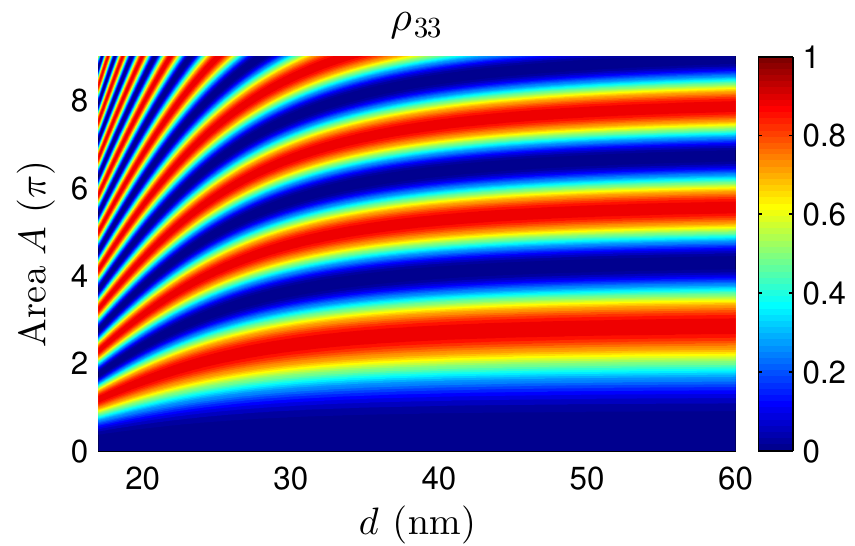}
\end{center}
\caption{\label{ContourPlotRho33-vs-dHybrid} TPRO in the SQD-MNP hybrid considered in this paper as a function of the interparticle distance $d$ and the pulse area $A$. Plotted is the biexciton state population $\rho_{33}$ for a hybrid calculated numerically for an incident pulse with $t_0 = 20/\Delta_B = (2/3)$~ps.}
\label{ContourPlotRho33-vs-d-Hybrid}
\end{figure}

Figure~\ref{PopulationsDynamicsShortPulseHybrid} shows an example of the population dynamics for a short pulse ($t_0 = 20/\Delta_B = 2/3~\mathrm{ps} \ll \gamma_{21}, \gamma_{32})$, when the adiabatic limit is valid (see Sec.~\ref{Results}). Obviously, the characteristics of the time dependence of the populations are similar to those for an isolated SQD (compare with Fig.~4), only the number of Rabi cycles is larger due to the field enhancement effect.

The population dynamics for a long incident pulse ($t_0 = 900/\Delta_B = 30$ ps), whose duration ($\approx 60$ ps) already is comparable with the spontaneous decay time of the biexciton state, $\gamma_{32}^{-1} = 120$ ps, is shown in Fig.~9. We observe that the population of the biexciton state decreases during the pulse action due to the spontaneous emission. Simultaneously, the one-exciton is incoherently populated which partly destroys the TPRO.

The results presented in Figs.~\ref{ContourPlotsRho33andRho22Hybrid}~-~\ref{PopulationsDynamicsLongPulseHybrid} were obtained for the SQD-MNP separation $d = 18$~nm. Figure~\ref{ContourPlotRho33-vs-dHybrid} shows how the TPRO depends in this separation. The contour plot was obtained numerically for an incident pulse with $t_0 = 20/\Delta_B = (2/3)$~ps. The figure clearly shows that the number of Rabi cycles decreases
with increasing $d$, which is due to the reduction of the enhancement factor $|1 + \alpha(\omega_0)/(2\pi d^3)|$ for growing $d$. As a result, the pulse areas $A$, at which the Rabi oscillations attain their successive maxima, grow with $d$.

\section{Effects of the SQD self-action}
\label{Self-action effect}
In this section, we discuss the effect of the SQD-MNP self-action described by the complex-valued feedback parameters $G_i$ ($i = 1, 2, 3$), Eqs.~(\ref{G1})~-~(\ref{G3}) on the TPRO of the hybrid. The feedback parameters are hidden in equations of motion~(\ref{rho11})~-~(\ref{rho31}). Their role is uncovered after substituting into Eqs.~(\ref{rho21}) and~(\ref{rho31}) the explicit expressions for the Rabi amplitudes $\Omega_{21}$ and $\Omega_{32}$, Eqs.~(\ref{Omega21}) and~(\ref{Omega32}). The principal effects of the feedback are most pronounced in Eqs.~(\ref{rho21}) and~(\ref{rho32}):
\begin{widetext}
\begin{subequations}
\begin{eqnarray}
\label{rho21extended}
\dot{\rho}_{21} = &-& \left[ i \left(\Delta_{21} + G_{1}^R Z_{21} \right)
                +\frac{1}{2}\gamma_{21} - G_{1}^I Z_{21} \right] \rho_{21}
\nonumber\\
               &+& i ( \widetilde{\Omega}^{0*}_{32} \rho_{31}
                - \widetilde{\Omega}^0_{21} Z_{21} )
                + i \left[ (G^*_3 \rho^*_{21} + G^*_2 \rho^*_{32}) \rho_{31}
                - G_3 \rho_{32} Z_{21} \right]~,
\end{eqnarray}
\begin{eqnarray}
\label{rho32extended}
\dot{\rho}_{32} = &-& \left[ i(\Delta_{32} + G_{2}^R Z_{32})
                + \frac{1}{2} (\gamma_{32} + \gamma_{21}) - G_{2}^I Z_{32} \right] \rho_{32}
\nonumber\\
                &-& i ( \widetilde{\Omega}^{0*}_{21} \rho_{31} + \widetilde{\Omega}^0_{32} Z_{32} )
                - i \left[ (G^*_1 \rho^*_{21} + G^*_3 \rho^*_{32}) \rho_{31} + G_3 \rho_{21} Z_{32} \right]~.
\end{eqnarray}
\end{subequations}
\end{widetext}
As follows from Eqs.~(\ref{rho21extended}) and~(\ref{rho32extended}), the self-action of the SQD gives rise to many additional nonlinearities as compared to an isolated SQD ($G_i = 0$). Two of these should be especially mentioned, namely: (i) - renormalization of the SQD frequencies, $\omega_2 \rightarrow \omega_2 + G_1^R Z_{21}$ and  $\omega_3 \rightarrow \omega_3 + G_2^R Z_{32}$, and (ii) - renormalization of the relaxation rates of the off-diagonal density matrix elements, $\gamma_{21}/2 \rightarrow \gamma_{21}/2 - G_1^I Z_{21}$ and $(\gamma_{21} + \gamma_{32})/2 \rightarrow (\gamma_{21} + \gamma_{32})/2 - G_2^I Z_{32}$ [compare the expressions in square brackets in Eqs.~(\ref{rho21}) and~(\ref{rho32})], both depending on the corresponding population differences. Thus, during a pulsed excitation, which gives rise to time-dependent population differences,
both the effective detunings away from resonance, $\Delta_{21} + G_{1}^R Z_{21}$ and $\Delta_{32} + G_{2}^R Z_{32}$,  and the effective relaxation rates of the transitions, $(1/2)\gamma_{21} - G_{1}^I Z_{21}$ and $(1/2) (\gamma_{32} + \gamma_{21}) - G_{2}^I Z_{32}$, will be swept during the pulse action, which does not occur in the case of an isolated SQD ($G_i=0$). These two effects may, in principle, substantially modify the plasmon-assisted TPRO.

In our case of the CdSe/ZnSe-Au hybrid, the magnitudes of all feedback parameters $\hbar|G_i|$, calculated for the minimal center-to-center spacing $d = 18$~nm we used in our computations, turn out to be on the order of a few tenths of meV, i.e. hundred times smaller than the biexciton binding energy $\hbar\Delta_B = 20$~meV. The spectral width of the longest pulse we employed ($t_0 = 900/\Delta_B = 30$~ps) is $\approx\hbar\Delta_B/1800 \approx 0.1$~meV, i.e. comparable with $|G_i|$, while for smaller $t_0$ it is even larger. Thus, the effects of the SQD self-action do not play any role for the range of parameters used to obtain the results presented in Sec.~\ref{Hybrid SQD}. Nevertheless, the SQD self-action effect might be of importance for other hybrids such as, for instance, a ZnSe/ZnS(core/shell)-Ag heterodimer for which $\hbar\Delta_B = 2.5$~meV and all $|G_i|$ (taken at $d = 16$~nm) are on the order of $\Delta_B$~\cite{NugrohoJOpt2017}.
This situation requires an additional study due to the complicated interplay of the enhancement and self-action effects.


\section{Summary}
\label{Summary}
We conducted a theoretical study of the two-photon Rabi oscillations of a heterodimer comprising a semiconductor quantum dot and a metal nanosphere,
considering the SQD as a three-level ladder-like system with ground, one-exciton and biexciton states. The Rabi oscillations in an isolated InGaAs/GaAs quantum dot have been investigated earlier in Ref.~\onlinecite{StuflerPRB2006}. We took into account the spontaneous decay of the excitonic states, which has not been done in Ref.~\onlinecite{StuflerPRB2006}, and found a tradeoff between the adiabatic regime, supporting coherent TPRO, and spontaneous processes. While in the former case, the TPRO can be realized, in the latter the spontaneous emission destroys the TPRO for the duration of incident pulses comparable to or longer than the spontaneous emission time. This limits the pulse width for which coherent TPRO can be observed.

The presence of a MNP nearby the SQD results in an enhancement of the external field magnitude depending on the SQD-MNP center-to-center distance, which leads to increasing the number of Rabi cycles per pulse as compared to an isolated SQD at the same magnitude of the external field. This effect may be advantageous for quantum technological applications that require the production of many entangled photon pairs per second.

We performed our calculations for a model system that may be realized in practice: a heterodimer comprised of a closely spaced CdSe/ZnSe quantum dot and a gold nanosphere.
Other candidates to observe the plasmon-assisted TPRO are ZnSe/ZnS(core/shell)-Ag hybrids and heterodimers comprised of an InGaAs/GaAs quantum dot and a triangular silver nanoparticle, absorbing in a wide spectral range, from the visible to the infrared~\cite{WuNRL2015}.

\acknowledgments
This work was supported by the Directorate General of Higher Education, Ministry of Research,
Technology and Higher Education of Indonesia (PKLN research grant, contract No.098/SP2H/LT/DRPM/2018).
B.S.N. acknowledges the University of Groningen for hospitality.


\begin{thebibliography}{99}
%
\bibitem{HayatSST2011} A. Hayat, A. Nevet, P. Ginzburg, and M. Orenstein,
    Semicond. Sci. Technol. \textbf{26}, 83001 (2011).
%
\bibitem{MaruoOL1997} S. Maruo, O. Nakamura, and S. Kawata,
    Opt. Lett. \textbf{22}, 132 (1997).
%
\bibitem{Baldacchini2016} T. Baldacchini (Ed.), {\it Three-Dimensional Microfabrication Using Two-Photon Polymerization},
    Elsevier, 2016.
%
\bibitem{SvobodaNeuron2006} K. Svoboda and R. Yasuda, Neuron \textbf{50}, 823 (2006)..
%
\bibitem{StricklerOptLett1991} J. H. Strickler and W. W. Webb,
    Opt. Lett. \textbf{16}, 1780 (1991).
%
\bibitem{CorredorAdvMat2006} C. C. Corredor, Z. L. Huang, K. D. Belfield,
    Adv. Mat. \textbf{18}, 2910 (2006).
%
\bibitem{MakarovJOSAB2007} N. S. Makarov, A. Rebane, M. Drobizhev, H. Wolleb, and H. Spahni, J. Opt. Soc. Am. B, \textbf{24}, 1874 (2007).
%
\bibitem{NeilsonBook2000} M. A. Neilson and I. L. Chuang,
    {\it Quantum Computation and Quantum Information} (Cambridge University Press, Cambridge, 2000).
%
\bibitem{MacchiavelloBook2000} C. Macchiavello, G. M. Palma, and A. Zeilinger,
    {\it Quantum Computation and Quantum Information Theory} (World Scientific, Singapore, 2000).
%
\bibitem{JenneweinPRL2000} T. Jennewein, C. Simon, G. Weihs, H. Weinfurter, and A. Zeilinger,
    Phys. Rev. Lett. \textbf{84}, 4729 (2000).
%
\bibitem{BensonPRL2000} O. Benson, C. Santori, M. Pelton, and Y. Yamamoto,
    Phys. Rev. Lett. \textbf{84}, 2513 (2000).
%
\bibitem{StacePRB2003} T. M. Stace, G. J. Milburn, and C. H. W. Barnes,
    Phys. Rev. B \textbf{67}, 085317 (2003).
%
\bibitem{FlissikowskiPRL2004} T. Flissikowski, A. Betke, I. A. Akimov, and F. Henneberger,
    Phys. Rev. Lett. \textbf{92}, 227401 (2004).
%
\bibitem{StevensonNature2006} R. M. Stevenson, R. J. Young, P. Atkinson, K. Cooper, D. A. Ritchie and A. J. Shields,
    Nature \textbf{439}, 179 (2006).
%
\bibitem{AkimovPRL2006} I. A. Akimov, J. T. Andrews, and F. Henneberger,
    Phys. Rev. Lett. \textbf{96}, 067401 (2006).
%
\bibitem{AkopianPRL2006} N. Akopian, N. H. Lindner, E. Poem, Y. Berlatzky, J. Avron, and D. Gershoni,
    Phys. Rev. Lett. \textbf{96}, 130501 (2006).
%
\bibitem{delValleNJP2011} E. del Valle, A. Gonzalez-Tudela, E. Cancellieri, F. P. Laussy and C. Tejedor,
    New J. Phys. \textbf{13}  113014 (2011).
%
\bibitem{SchumacherOE2012} S. Schumacher, J. F\"orstner, A. Zrenner, M. Florian, C. Gies, P. Gartner, and F. Jahnke.
    Opt. Express \textbf{20}, 5335 (2012).
%
\bibitem{delVallePRB2010} E. del Valle, S. Zippilli, F. P. Laussy, A Gonzalez-Tudela, G. Morigi, and C. Tejedor,
    Phys. Rev. B \textbf{81}, 035302 (2010).
%
\bibitem{StuflerPRB2006} S. Stufler, P. Machnikowski, P. Ester, M. Bichler, V. M. Axt, T. Kuhn, and A. Zrenner,
     Phys. Rev. B \textbf{73}, 125304 (2006).
%
\bibitem{BorriPRB2002} P. Borri, W. Langbein, S. Schneider, U. Woggon, R. L. Sellin, D. Ouyang, and D. Bimberg,
    Phys. Rew. B \textbf{66}, 081306(R), (2002).
%
\bibitem{ChenPRL2002} G. Chen, T. H. Stievater, E. T. Batteh, X. Li, D. G. Steel, D. Gammon, D. S. Katzer, D. Park, and L. J. Sham,
   Phys. Rev. Lett. \textbf{88}, 117901 (2002).
%
%
\bibitem{JayakumarPRL2013} H. Jayakumar, A. Predojevi\'c, T. Huber, T. Kauten, G. S. Solomon, and G. Weihs,
    Phys. Rev. Lett. \textbf{110}, 135505 (2013).
 %
\bibitem{MachnikowskiPRB2008} P. Machnikowski,
     Phys. Rev. B \textbf{78}, 195320 (2008).
%
\bibitem{ArtusoNL2008} R. D. Artuso and G. W. Bryant,
    Nano Lett. \textbf{8}, 2106 (2008).
%
\bibitem{ArtusoPRB2010} R. D. Artuso and G. W. Bryant,
    Phys. Rev. B \textbf{82}, 195419 (2010).
%
\bibitem{MalyshevPRB2011} A. V. Malyshev and V. A. Malyshev,
    Phys. Rev. B \textbf{84}, 035314 (2011).
%
\bibitem{LiOE2012} J. B. Li, N. C. Kim, M. T. Cheng, L. Zhou, Z. H. Hao, and Q. Q Wang,
    Opt. Express \textbf{20}, 1856 (2012).
%
\bibitem{NugrohoJCP2013} B. S. Nugroho, A. A. Iskandar, V. A. Malyshev, and J. Knoester,
    J. Chem. Phys. \textbf{139}, 014303 (2013).
%
\bibitem{ZhangPRL2006} W. Zhang, A. O. Govorov, and G. W. Bryant,
    Phys. Rev. Lett. \textbf{97}, 146804 (2006).
%
\bibitem{KosionisJPCC2012} S. Kosionis, A. Terzis, V. Yannopapas, and E. Paspalakis,
    J. Phys. Chem. C \textbf{116}, 23663 (2012).
%
\bibitem{NugrohoPRB2015} B. S. Nugroho, A. A. Iskandar, V. A. Malyshev, J. Knoester,
    Phys. Rev. B \textbf{}, (2015).
%
\bibitem{SadeghiNanotechnology2010} S. M. Sadeghi,
    Nanotechnology \textbf{21}, 455401 (2010).
%
\bibitem{SadeghiPRB2009} S. M. Sadeghi,
    Phys. Rev. B \textbf{79}, 233309 (2009).
%
\bibitem{AntonPRB2012} M. Ant\'on, F. Carre\~no, S. Melle, O. Calder\'on, E. Cabrera-Granado, J. Cox, and M. Singh,
    Phys. Rev. B \textbf{86}, 155305 (2012).
%
\bibitem{NugrohoJOpt2017} B. S. Nugroho, A. A. Iskandar, V. A. Malyshev, J. Knoester, J. Opt. \textbf{19}, 015004 (2017).
%
\bibitem{MessiahBook1966} A. Messiah, {\it Quantum Mechanics} (North-Holland, Amsterdam, 1966).
%
\bibitem{Bohren1983} C. F. Bohren and  D. R. Huffman, {\it Absorption and
    Scattering of Light by Small Particles}; Wiley: New York, 1983.
%
\bibitem{Maier2007} S. A. Maier,  {\em Plasmonics: Fundamentals and
    Applications}, Springer: New York, 2007.
%
\bibitem{SchollNature2012} J. A. Scholl, A. L. Koh, and J. A. Dionne, Nature {\bf 483}, 421 (2012).
%
\bibitem{JundtPRL2008} G. Jundt, L. Robledo, A. H\"ogele, S. F\"alt, and A. Imamo\u{g}lu,
    Phys. Rev. Lett. \textbf{100}, 177401 (2008).
%
\bibitem{GerardotNJP2009} B. Gerardot, D. Brunner, P. Dalgarno, K. Karrai, A. Badolato, P. Petroff, and R. Warburton,
     New J. Phys. \textbf{11}, 013028 (2009).
%
\bibitem{YanPRB2008} J. Y. Yan, W. Zhang, S. Duan, X. G. Zhao, and A. O. Govorov,
    Phys. Rev. B {\bf 77}, 165301 (2008).
%
%
%
%
\bibitem{Lindblad1976} G. Lindblad, Commun. Math. Phys. {\bf 48}, 119 (1976).
%
\bibitem{Blum2012} K. Blum, {\it Density Matrix: Theory and applications}, 3rd edition (Springer, 2012).
%
%
\bibitem{DerkachovaPlasmonics2016} A. Derkachova, K. Kolwas, and I. Demchenko, Plasmonics \textbf{11}, 941 (2016).
%
%
\bibitem{WuNRL2015} C. Wu, X. Zhou, and J. Wei, Nanoscale Res. Lett. \textbf{10}, 354 (2015).
%
\end{thebibliography}

\end{document}